\documentclass[12pt,preprint]{aastex}

\def\kms{\ifmmode{\rm km\th s^{-1}}\else km\th s$^{-1}$\fi}
\def\th{\thinspace}

\shortauthors{Boden et al.}
\shorttitle{HD~98800~B}

\begin{document}

\title{Dynamical Masses for Low-Mass Pre-Main Sequence Stars: A
Preliminary Physical Orbit for HD~98800~B}

\email{**astro-ph version \today**}

\author{Andrew F.\ Boden\altaffilmark{1,2},
        Anneila I.\ Sargent\altaffilmark{3},
        Rachel L.\ Akeson\altaffilmark{1},
        John M.\ Carpenter\altaffilmark{3},
        Guillermo Torres\altaffilmark{4},
	David W.\ Latham\altaffilmark{4},
	David R.\ Soderblom\altaffilmark{5},
	Ed\ Nelan\altaffilmark{5},
	Otto G.\ Franz\altaffilmark{6},
	Lawrence H.\ Wasserman\altaffilmark{6}
}

\altaffiltext{1}{Michelson Science Center, California
Institute of Technology, 770 South Wilson Ave., Pasadena CA 91125}

\altaffiltext{2}{Department of Physics and Astronomy, Georgia State
University, 29 Peachtree Center Ave., Science Annex, Suite 400,
Atlanta GA 30303}

\altaffiltext{3}{Division of Physics, Math, and Astronomy, California
Institute of Technology, MS 105-24, Pasadena, CA 91125}

\altaffiltext{4}{Harvard-Smithsonian Center for Astrophysics, 60
Garden St., Cambridge MA 02138}

\altaffiltext{5}{Space Telescope Science Institute, 3700 San Martin
Dr, Baltimore, MD 21218}

\altaffiltext{6}{Lowell Observatory, 1400 W. Mars Hille Road, Flagstaff, AZ 86001}


\email{bode@ipac.caltech.edu}

\begin{abstract}

We report on Keck Interferometer observations of the double-lined
binary (B) component of the quadruple pre-main sequence (PMS) system
HD~98800.  With these interferometric observations combined with
astrometric measurements made by the Hubble Space Telescope Fine
Guidance Sensors (FGS), and published radial velocity observations we
have estimated preliminary visual and physical orbits of the HD~98800
B subsystem.  Our orbit model calls for an inclination of 66.8 $\pm$
3.2 deg, and allows us to infer the masses and luminosities of the
individual components.  In particular we find component masses of
0.699 $\pm$ 0.064 and 0.582 $\pm$ 0.051 M$_{\sun}$ for the Ba
(primary) and Bb (secondary) components respectively.

Spectral energy distribution (SED) modeling of the B subsystem
suggests that the B circumstellar material is a source of extinction
along the line of sight to the B components.  This seems to corroborate
a conjecture by Tokovinin that the B subsystem is viewed through
circumbinary material, but it raises important questions about the
morphology of that circumbinary material.

Our modeling of the subsystem component SEDs finds temperatures and
luminosities in agreement with previous studies, and coupled with the
component mass estimates allows for comparison with PMS models in the
low-mass regime with few empirical constraints.  Solar abundance
models seem to under-predict the inferred component temperatures and
luminosities, while assuming slightly sub-solar abundances bring the
models and observations into better agreement.  The present
preliminary orbit does not yet place significant constraints on
existing pre-main sequence stellar models, but prospects for
additional observations improving the orbit model and component
parameters are very good.

\end{abstract}


\section{Introduction}
\label{sec:introduction}

Accurate determinations of the physical properties of stars
(e.g.~mass, radius, temperature, luminosity, elemental abundance,
etc.) provide fundamental tests of stellar structure and evolution
models.  The most basic of stellar physical properties is the mass,
available only from the study of dynamical interactions -- most
typically in binary stars.

Among the areas where our understanding of stellar structure is most
uncertain is in pre-main sequence (PMS) stars, particularly for
low-mass systems \citep[see][and references for
summaries]{Palla2001,Hillenbrand2004}.  Providing empirical
constraints on models of PMS stars is critical for improving their
accuracy, and the reliability of these models in turn is critical to
our understanding of individual PMS systems in particular, and the
process of star formation in general.  So experimental determinations
of masses and luminosities for PMS stars are of fundamental importance
in constraining our understanding of star formation and early stellar
evolution.

Eclipsing binary systems are important because of the favorable
geometry for accurate mass and radius determination
\citep[e.g.][]{Andersen1991}.  However, component mass determinations
for non-eclipsing spectroscopic binaries are possible with measurement
of the orbital inclination.  In order to expand the number of low-mass
PMS systems with dynamical mass measurements, we have embarked on a
program to estimate physical orbits in PMS systems by integrating
interferometric and spectroscopic observations.  This follows the
initial determination of a non-eclipsing PMS system orbit for the
T~Tauri binary HBC~427 by \citet{Steffen2001}.  Additionally
\citet{Simon2000} have used star-disk interactions to measure the mass
of PMS stars.  \cite{Hillenbrand2004} provide a current summary of
dynamical mass determinations for pre-main sequence stars.

\objectname[HD 98800]{HD~98800} (HIP 55505,
TWA 4A) is a well-studied quadruple star system in the TW Hya
association.  The system was first detected as a visual binary by
\citet{Innes1909}; at current epoch the separation is roughly 0.8'' on
approximately a N-S line \citep[herein P2001]{Prato2001}.
\citet[][herein T95]{Torres1995} established that both visual
components are themselves spectroscopic binaries, finding a 262-day
single-lined orbit for the primary (Southern, A) component, and a
315-day double-lined orbit for the secondary (Northern, B) component.
The system exhibits a strong mid-infrared (IR) excess longward of 7
$\mu$m \citep{Walker1988,Zuckerman1993}, lithium signatures, but no
signs of active accretion \citep{Soderblom1996,Webb1999}.  The mid-IR
excess, Li, HD~98800's putative membership in the TW Hya association,
and the Hipparcos distance estimate of 46.7 $\pm$ 6.2 pc (establishing
component luminosities) lead to the consensus that the system is PMS.
\citet[herein S98]{Soderblom1998} has termed the system ``post-T
Tauri'', and estimates a system age range of 5 -- 20 Myr with a most
likely value of 10 Myr based primarily on Li abundance.  Multi-band IR
imaging studies firmly establish that the mid-IR excess is associated
with the double-lined B subsystem, most likely in the form of a
circumbinary disk \citep[see][ P2001, and references
therein]{Koerner2000}.

Here we report on observations of the double-lined HD~98800~B binary
subsystem made with the Keck Interferometer
\citep[KI,][]{Colavita2003} and the Hubble Space Telescope Fine
Guidance Sensor (FGS).  These observations resolve the B subsystem and
allow us to estimate the visual and physical orbits (in combination
with radial velocity measurement from T95), and determine the
component dynamical masses and luminosities.  KI and FGS observations
of HD~98800~B and our orbital solution are described in
\S~\ref{sec:observations}.  Physical parameters implied by the orbit
and spectral energy distribution modeling are discussed in
\S~\ref{sec:physics}.  Finally a discussion of these results including
comparisons with pre-main sequence stellar models is given in
\S~\ref{sec:discussion}.

\section{Observations and Orbital Solution}
\label{sec:observations}

\paragraph{KI Observations}
The KI interferometric observable used for these measurements is the
fringe contrast or {\em visibility} (squared) of an observed
brightness distribution on the sky.  KI was used to make the
interferometric measurements presented here; KI is a long-baseline $H$
(1.6$\mu$m) and $K$-band (2.2$\mu$m) interferometer located at Mauna
Kea, HI, and described in detail elsewhere \citep{Colavita2003}.  The
analysis of such data on a binary system is discussed in detail in
previous work \citep[e.g.][]{Boden2000,Hummel2001}.

HD~98800~B was observed in conjunction with calibration objects by KI
in $K$-band ($\lambda \sim 2.2 \mu$m) on five nights between 18 April
2003 and 22 April 2005, a dataset spanning roughly two years and 2.3
orbital periods.  HD~98800~B and relevant calibration objects were
observed multiple times during each of these nights, and each
observation (scan) was approximately 130 sec long.  For each scan we
computed a mean $V^2$ value from the scan data, and the error in the
$V^2$ estimate is inferred from the rms internal scatter.  HD~98800~B
was always observed in combination with one or more calibration
sources within $\sim$ 5$^{\circ}$ on the sky.  For this analysis we
have used \objectname[HD 97590]{HD~97590} (A6 V) and \objectname[HD
100219]{HD~100219} (F7 V) as calibration objects; Table
\ref{tab:calibrators} lists the relevant observational parameters for
the calibration objects.  Calibrating our interferometric data with
respect to these objects results in 34 calibrated visibility scans on
HD~98800~B; these measurements are summarized in Table
\ref{tab:V2Table}.  Finally we will note that as the Keck Telescopes
separately resolve the HD~98800 A-B system (P2001), and the KI beam
combiner is fed by single-mode fiber, no light from HD~98800~A falls
on the fringe camera when the device is measuring HD~98800~B.
Consequently no special provisions are necessary in processing KI
observations of HD~98800~B.

\begin{deluxetable}{cccccc}
\tabletypesize{\small}
\tablecolumns{6}
\tablewidth{0pc}

\tablecaption{KI $V^2$ Calibration Objects Considered in our Analysis.
The relevant parameters for our KI calibration object are summarized.
Apparent diameter values are estimated from spectral energy
distribution modeling based on archival photometry and spectral energy
distribution templates from \cite{Pickles1998}.
\label{tab:calibrators}
}

\tablehead{
\colhead{Object} & \colhead{Spectral} & \colhead{$V$} & \colhead{$K$} & \colhead{HD~98800}   & \colhead{Adopted Model} \\
\colhead{Name}   & \colhead{Type}     &               &               & \colhead{Separation} & \colhead{Diameter (mas)}
}

\startdata
HD 97590    & A6 V     & 7.3 &  6.7 & 2.6$^{\circ}$  & 0.14 $\pm$ 0.04   \\
HD 100219   & F7 V     & 6.2 &  4.9 & 4.5$^{\circ}$  & 0.44 $\pm$ 0.03   \\
\enddata

\end{deluxetable}

\begin{table}
\dummytable\label{tab:V2Table}
\end{table}

\paragraph{HST FGS Observations}
The HD~98800 system was observed by the Hubble Space Telescope Fine
Guidance Sensors (FGS) in its ``FGS-TRANS'' mode and F583W filter on
20 epochs between 1996 and 2002.  Of course, the FGS are also
interferometers, but unlike the KI observations the FGS-TRANS data
have been processed into estimated Ba-Bb separations by methods
described in \citet{Franz1998}.  HD~98800 represents a challenging
target for FGS observation: first because the B subsystem separation
is near the resolution limit of the FGS, and second because (unlike
the KI observations) the visual A component flux must be accounted for
in data reduction.  Because of these difficulties, of these 20 epochs
only 11 of the measurements were deemed viable for triple-star
analysis.  These 11 B subsystem separation estimates are summarized in
Table \ref{tab:FGSdata}, which lists FGS separation data, model
predictions, and data - model residuals relative to our ``Joint-Fit''
orbital solution (Table \ref{tab:orbit}).

\begin{table}
\dummytable\label{tab:FGSdata}
\end{table}



\begin{figure}[p]
\epsscale{0.6}
\plotone{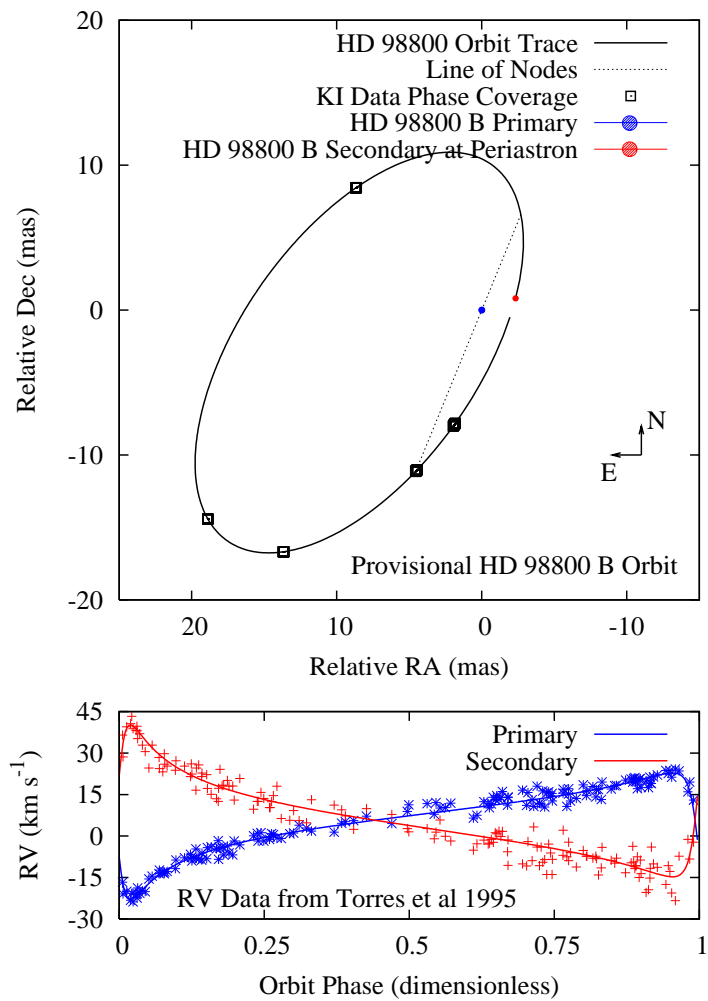}
\caption{Orbit of HD~98800~B as derived in our ``Joint-Fit'' solution
(Table~\ref{tab:orbit}).  Upper Panel: the relative visual orbit model
of HD~98800~B is shown, with the primary and secondary objects
rendered at T$_0$ (periastron).  The specific epochs where we have KI
V$^2$ phase coverage are indicated on the relative orbit (they are not
separation vector estimates).  Component diameter values are estimated
and rendered to scale.  Lower Panel: the double-lined radial velocity
orbit model and data from \citet{Torres1995}.
\label{fig:hd98800_orbit}}
\end{figure}

\paragraph{Orbital Solution}
In order to estimate the visual and physical orbit of HD~98800~B we
have integrated the astrometric datasets described above with the
double-lined radial velocity (RV) data on B presented in T95 Table 1.
Figure \ref{fig:hd98800_orbit} depicts our relative visual and
spectroscopic orbit model of the HD~98800~B subsystem as derived in
our ``Joint-Fit'' orbital solution (Table~\ref{tab:orbit}).  The upper
panel depicts the relative visual orbit model, with the primary (Ba)
component rendered at the origin, and the secondary (Bb) component
rendered at periastron.  We have indicated the phase coverage of our
KI V$^2$ data on the relative orbit with points (they are {\em not}
separation vectors); the phase coverage of the V$^2$ data is sparse
relative to other similar analyses \citep[e.g.][]{Boden2000}.  Because
of this sparse phase coverage of the V$^2$ data, our initial visual
orbit determination we constrained orbital parameters measured by RV
(i.e. $e$, $\omega$, $P$) to their T95 values (Table~\ref{tab:orbit}),
and found an orbit model that phased-up acceptably with the T95
solution.  Based on that constrained initial estimate, we then fully
integrated the V$^2$, FGS, and T95 RV data as described below
(Table~\ref{tab:orbit}).  The size of the HD~98800~B components are
estimated (\S~\ref{sec:SED}) and rendered to scale.  The lower panel
depicts the integrated double-lined spectroscopic orbit model and
radial velocity data from T95.

\begin{figure}[th]
\includegraphics[angle=-90,width=16cm]{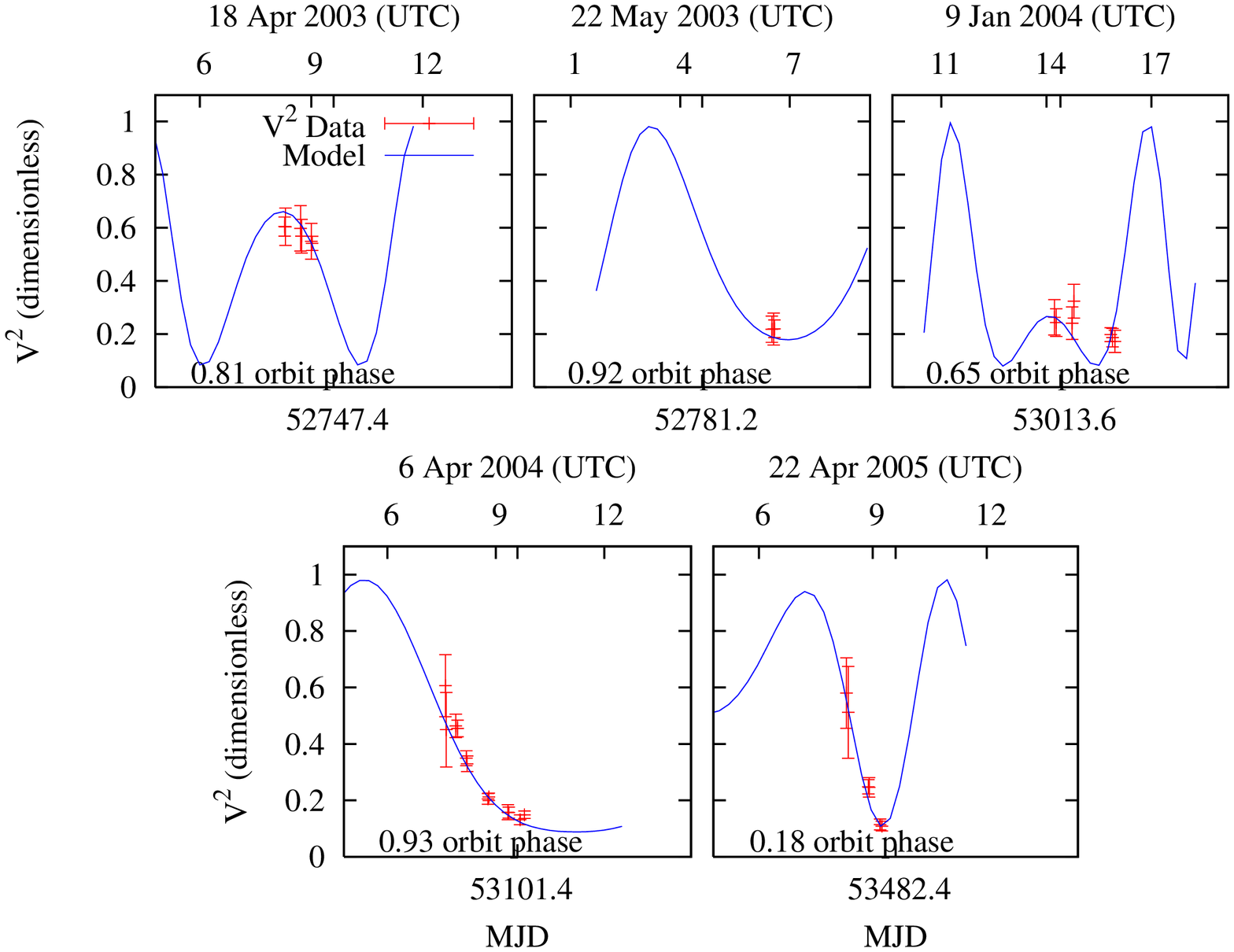}
\caption{KI Data/Model Comparisons for HD~98800~B V$^2$ Observations.
Here we give comparisons for our five epochs of KI V$^2$ observations
(Table \ref{tab:V2Table}).  In each case the data and model are shown.
\label{fig:dataComparisons}}
\end{figure}

\begin{figure}[th]
\plotone{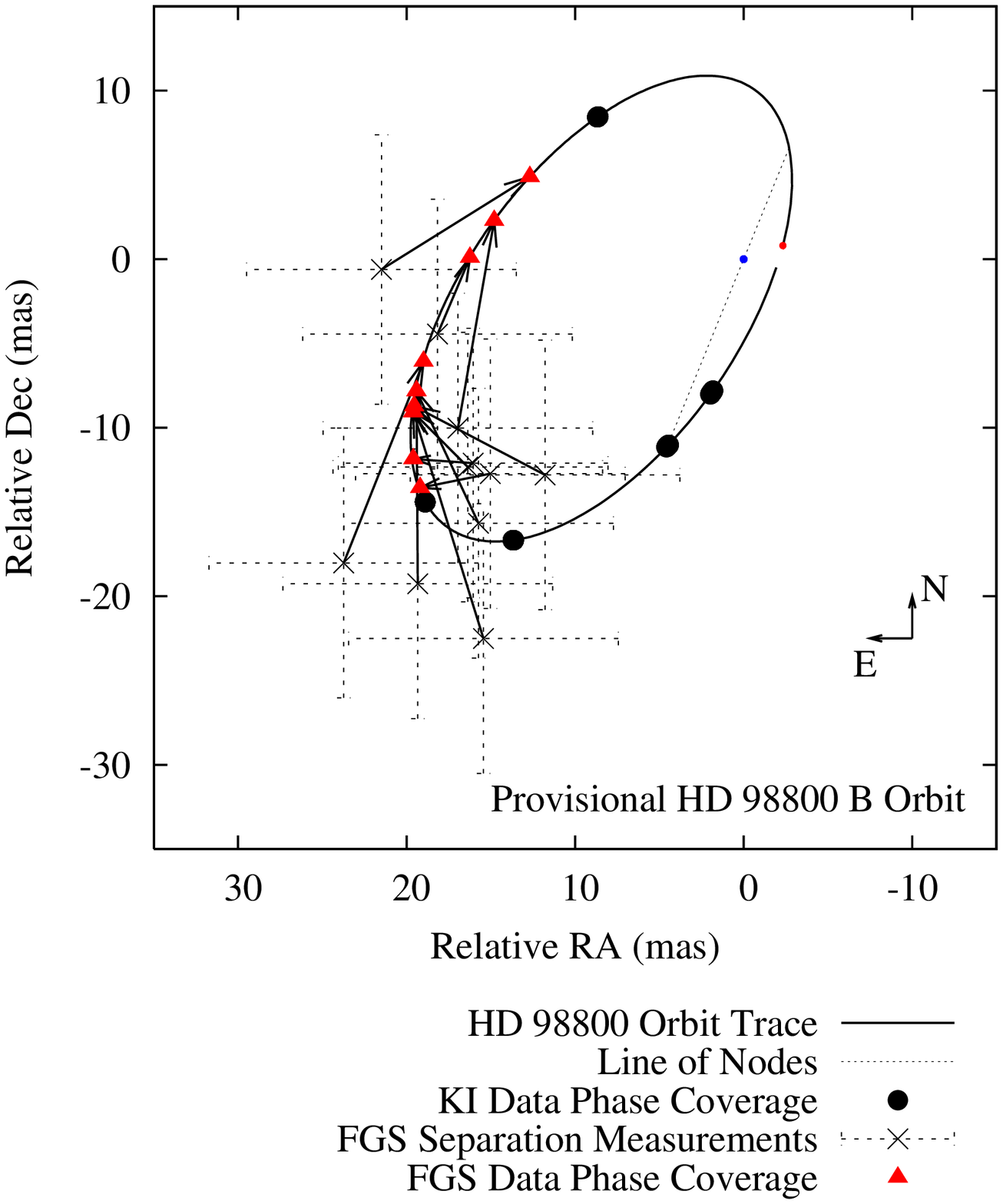}
\caption{FGS Data/Visual Orbit Model Comparisons for HD~98800~B
Observations.  Here we give comparisons for our 11 epochs of reduced
FGS observations (Table \ref{tab:FGSdata}) and the ``Joint-Fit''
visual orbit model from Table \ref{tab:orbit}.  Arrows indicate the
mapping of individual FGS separation estimates to corresponding points
(phases) on the visual orbit model.
\label{fig:FGSdata}}
\end{figure}

Figure \ref{fig:dataComparisons} depicts direct comparisons between
our KI V$^2$ observations and predictions from our HD~98800~B
``Joint-Fit'' orbit model (Table \ref{tab:orbit}; the five nights of
data are each rendered in separate subpanels).  The model is seen to be
in good agreement with the KI data.  Further, Figure \ref{fig:FGSdata}
shows a direct comparison between the FGS separation data and the
``Joint-Fit'' visual orbit model.  Again, the body of the FGS data are
in good agreement with our orbit model, and the FGS phase coverage
complements the phase coverage provided by the KI V$^2$ data.

\begin{figure}[p]
\includegraphics[angle=-90,width=8.5cm]{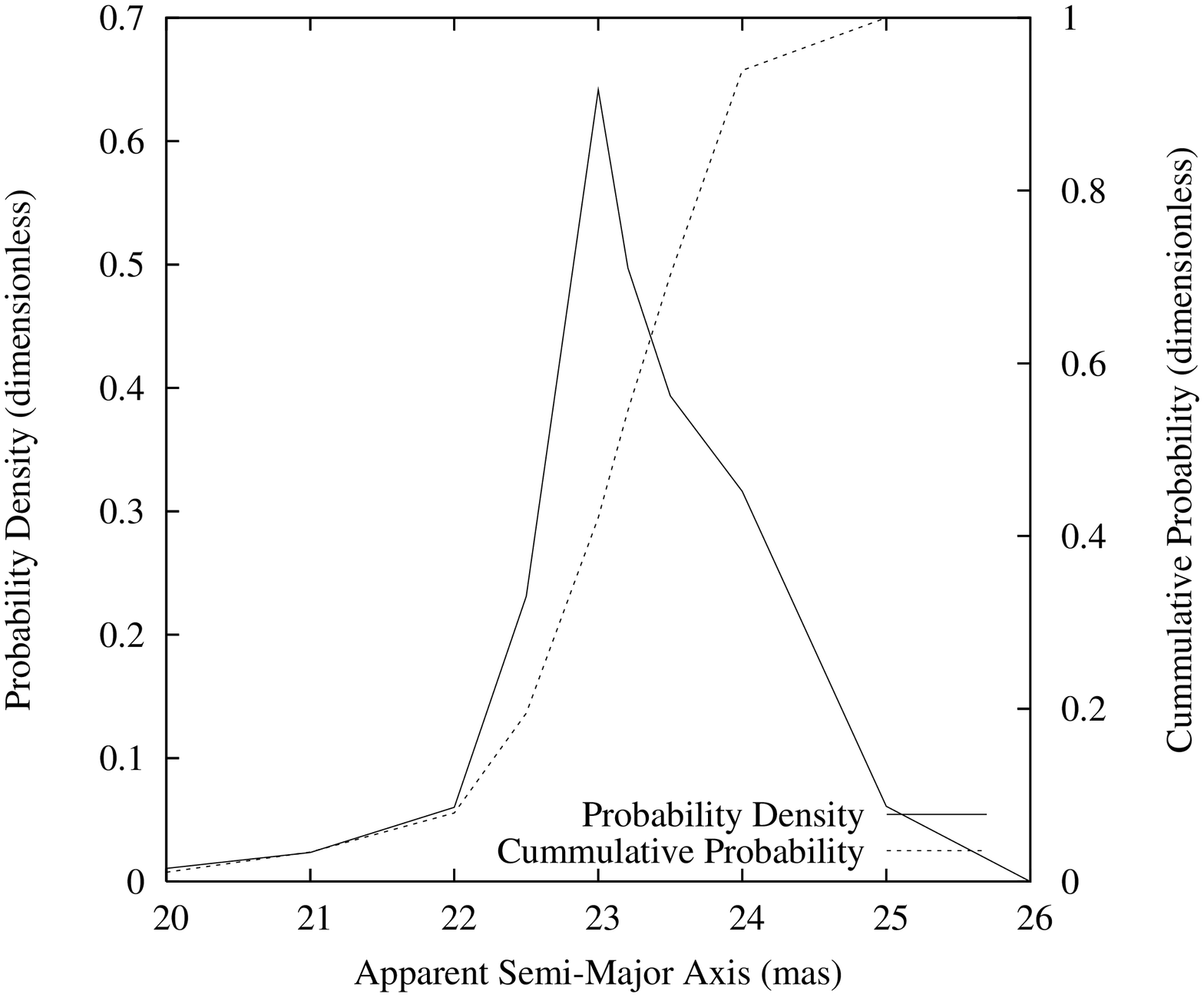}
\includegraphics[angle=-90,width=8.5cm]{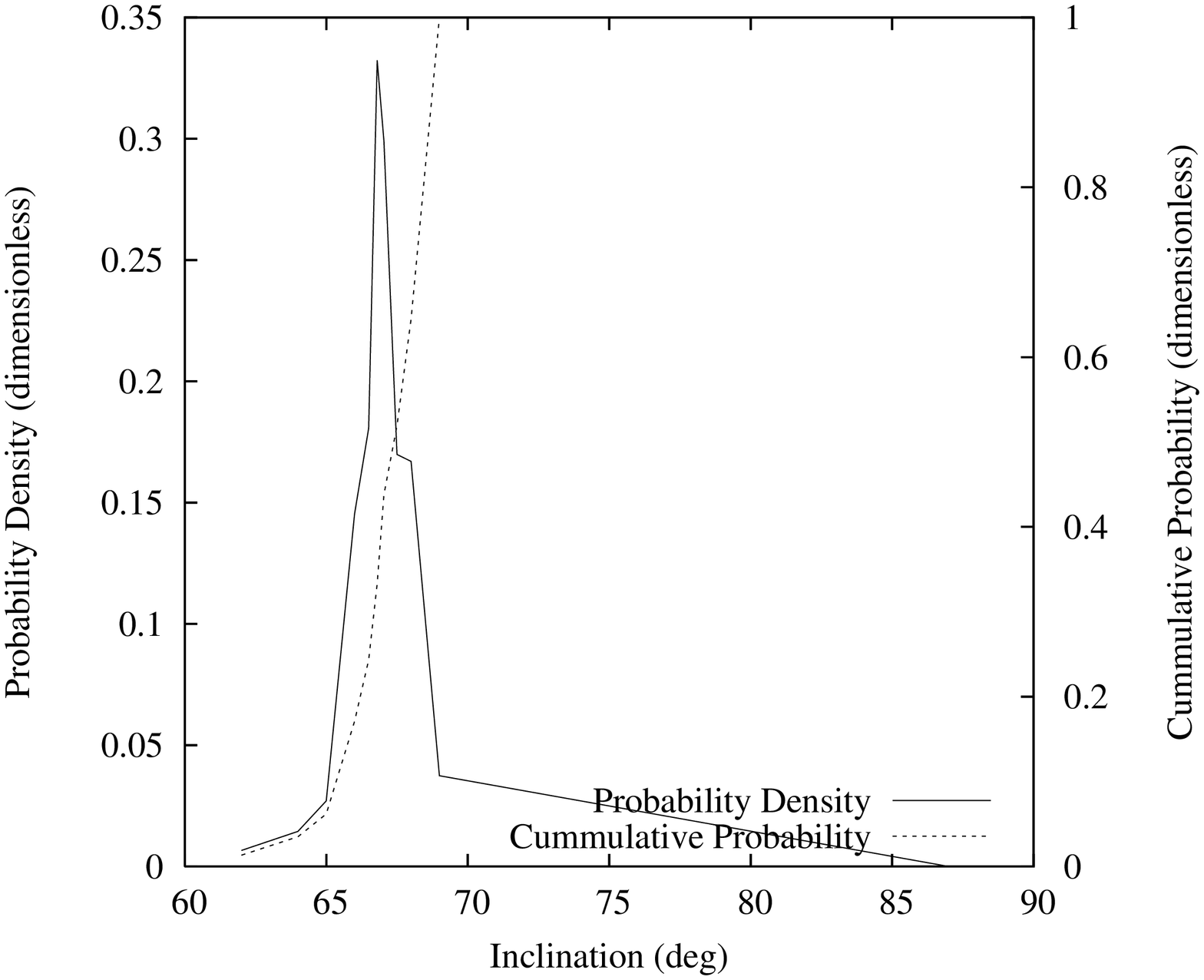}\\
\includegraphics[angle=-90,width=8.5cm]{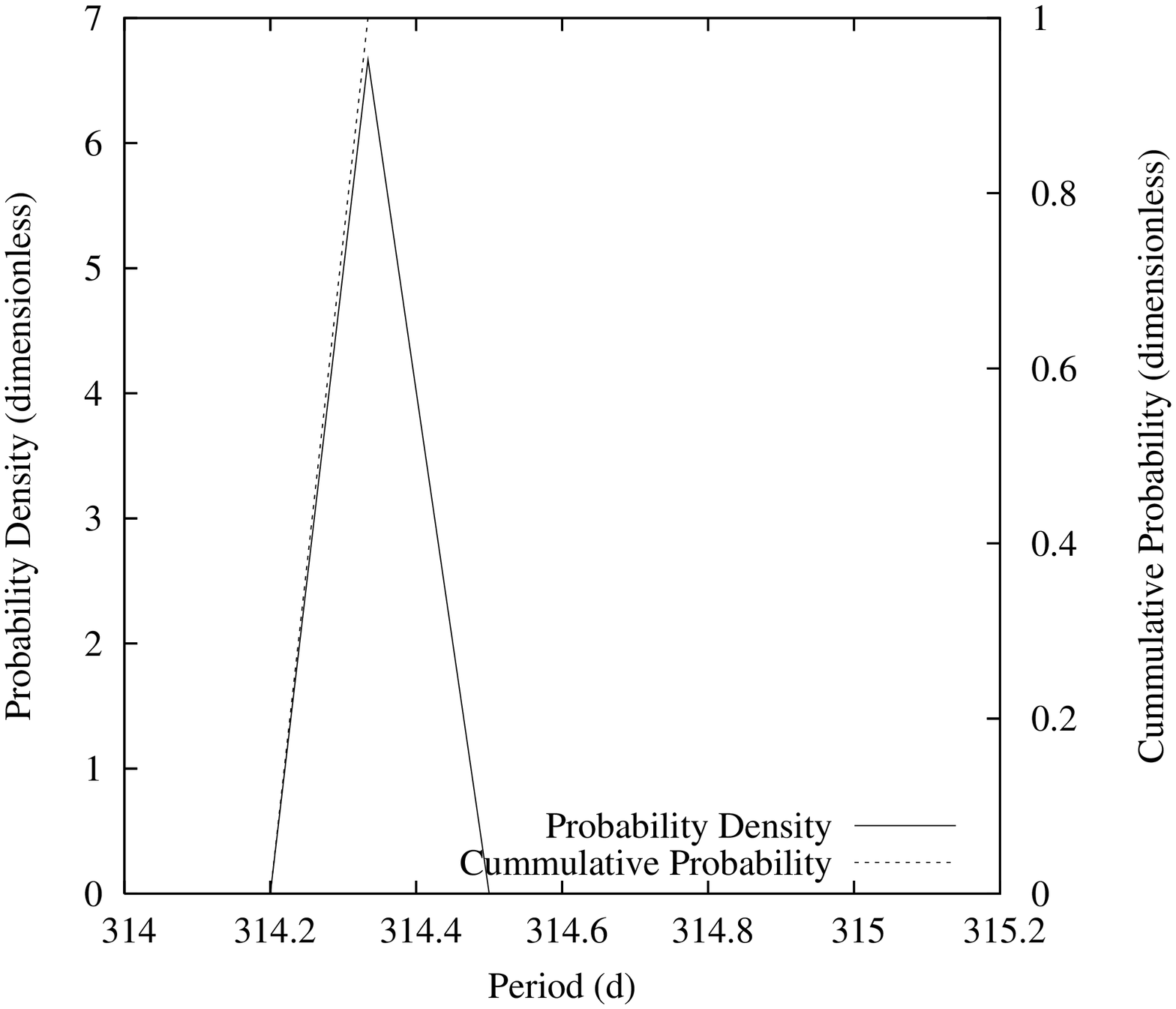}
\includegraphics[angle=-90,width=8.5cm]{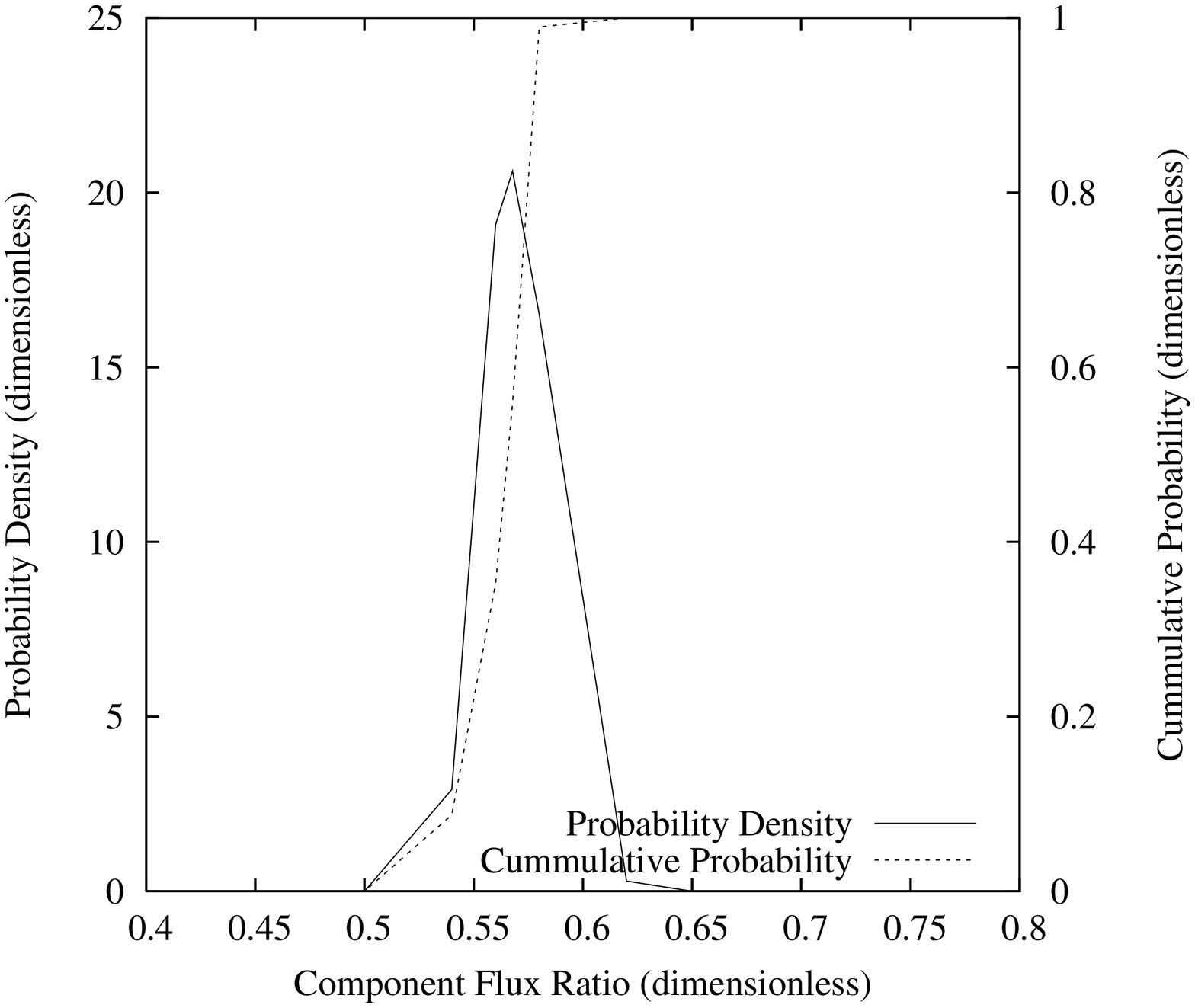}\\
\caption{Bayesian Orbital Parameter Probability Density Estimates.  To
evaluate the uncertainties in our orbital model we conducted
independent Marquardt-Levenberg least-squares and Bayesian parameter
estimation analyses on the integrated KI V$^2$/FGS/T95 RV dataset.
Shown here are probability density and cumulative probability
parameter estimates from the Bayesian analysis for four important
orbital parameters (clockwise from top left: apparent semi-major axis
($a$), inclination ($i$), component intensity ratio, and period
($P$)).  Agreement in orbital parameter values and uncertainties
between the two independent analyses is good.
\label{fig:BayesianDist}}
\end{figure}

In order to eliminate the possibility of multiple orbital solutions
consistent with the relatively sparse KI data, we augmented our
traditional Marquardt-Levenberg least-squares analysis
\citep[e.g.][]{Boden2000} with a Bayesian parameter estimation
analysis \citep{Bretthorst1988,Press1992,Lay1997,Akeson2002} using the
integrated KI/FGS/RV dataset, an assumption of uniform priors in the
orbital parameters, and a standard data likelihood function based on
the chi-squared agreement between the observation set and orbit model:
\begin{displaymath}
P(D|Model) \propto \exp \left[ - \sum_{k}
                          \frac{(d_k - {\hat d}_k(Model))^2}
			       {2 \sigma_{k}^2} \right]
\end{displaymath}
Figure \ref{fig:BayesianDist} shows representative parameter estimate
probability density distributions from this process.
The Bayesian analysis indicated no viable orbital solutions
beyond the one presented here, and the orbital parameter estimates and
uncertainties produced by the Marquardt-Levenberg and Bayesian
analyses are in good agreement.

Orbital parameter estimates for the 98800~B subsystem are summarized
in Table~\ref{tab:orbit}.  Included for comparison are the
double-lined orbital parameters on B from T95 (``T95''), the visual
orbit parameters estimated holding $P$, $e$, and $\omega$ to their T95
values (``V$^2$ Constrained''), and solutions integrating the KI V$^2$
and double-lined RV data from T95 (``V$^2$'') and V$^2$, FGS
separations, and T95 RV (``Joint-Fit'') .  The various orbital
solutions are in good agreement; the most notable variation ($\approx$
2$\sigma$) appears between the T95 and Joint-Fit period estimates
based on the expanded time baseline of the combined dataset.  In both
the Marquardt-Levenberg and Bayesian orbital analysis we have weighted
the astrometric datasets (KI and FGS) and T95 RV data equally and so
as to yield a solution chi-squared per degree of freedom of 1: the
resulting V$^2$ residual statistics are consistent with previous
analyses from KI \citep{Colavita2003}, and the FGS residual statistics
are in reasonable agreement with previous FGS analyses given the
complexity of the HD~98800 FGS data
\cite[e.g.][]{Franz1998,Benedict2001}.  Uncertainties quoted in Tables
\ref{tab:V2Table} and \ref{tab:FGSdata} reflect the weightings in the
Joint-Fit solution.

\begin{deluxetable}{l|c|ccc}
\tablecolumns{5}
\tablewidth{0pc}

\tablecaption{Orbital Parameters for HD~98800~B.  Summarized here are
orbital parameters for the HD~98800~B system as determined by T95 and
present results.  We give three separate fits to our KI and FGS data:
V$^2$ only (constrained in $P$, $e$, and $\omega$ to T95 values as
indicated by brackets, ``V$^2$ Constrained''), KI V$^2$ integrated with
T95 radial velocities (``V$^2$ \& RV''), and KI V$^2$, FGS, and T95 RV
(``Joint-Fit'') .  Note that the V-band Bb/Ba component flux ratio has
been re-estimated based on the original T95 spectra; the revised value
is given here.
\label{tab:orbit}
}

\tablehead{
\colhead{Orbital}        & \colhead{T95}        & \multicolumn{3}{c}{This Work} \\
\colhead{Parameter}      &                      & \colhead{V$^2$ Constrained} & \colhead{V$^2$ \& RV}   & \colhead{Joint-Fit}
}
\startdata
Period (d)               & 315.15 $\pm$ 0.39    & [315.15]                    & 314.334 $\pm$ 0.029   &  314.327  $\pm$ 0.028 \\
T$_{0}$ (MJD)            & 48709.48 $\pm$ 0.30  & 52479.74 $\pm$ 0.38         & 52481.32 $\pm$ 0.23   &  52481.34 $\pm$ 0.22 \\
$e$                      & 0.7812 $\pm$ 0.0059  & [0.7812]                    & 0.7836 $\pm$ 0.0053   &  0.7849 $\pm$ 0.0053 \\
K$_{Ba}$ (km s$^{-1}$)   & 22.54 $\pm$ 0.37     &                             & 22.87  $\pm$ 0.34     &  22.94 $\pm$ 0.34    \\
K$_{Bb}$ (km s$^{-1}$)   & 27.04 $\pm$ 0.62     &                             & 27.45  $\pm$ 0.60     &  27.53 $\pm$ 0.61    \\
$\gamma$ (km s$^{-1}$)   & 5.73 $\pm$ 0.14      &                             & 5.72   $\pm$ 0.14     &  5.73  $\pm$ 0.14    \\
$\omega_{Ba}$ (deg)      & 109.5 $\pm$ 1.2      & [109.5]                     & 108.9  $\pm$ 1.1      &  109.6 $\pm$ 1.1     \\
$\Omega$ (deg)           &                      & 337.4 $\pm$ 3.2             & 337.8  $\pm$ 2.9      &  337.6 $\pm$ 2.4     \\
$i$ (deg)                &                      & 66.7 $\pm$ 3.9              & 67.0   $\pm$ 3.7      &  66.8  $\pm$ 3.2     \\
$a$ (mas)                &                      & 23.0 $\pm$ 3.5              & 23.2   $\pm$ 3.1      &  23.3  $\pm$ 2.5     \\
$\Delta K$ (mag)          &                      & 0.64 $\pm$ 0.18            & 0.616  $\pm$ 0.050    &  0.612 $\pm$ 0.046   \\
$\Delta V$ (mag)         & {\em 1.11  $\pm$ 0.15}   &                         &                       &     \\
\hline
\enddata

\end{deluxetable}

\section{Physical properties of HD~98800~B}
\label{sec:physics}

The orbital parameters from Table~\ref{tab:orbit} allow us to directly
compute many of the physical properties of the HD~98800~B subsystem
and its components.  Physical parameters derived from our HD~98800~B
``Joint-Fit'' integrated visual/spectroscopic orbit are summarized in
Table \ref{tab:physics}.  The limited V$^2$ phase coverage and
resulting modest-precision orbital solution (Table \ref{tab:orbit})
yields preliminary dynamical masses of 0.699 $\pm$ 0.064 and 0.582
$\pm$ 0.051 M$_{\sun}$ for the primary (Ba) and secondary (Bb)
components respectively.

The Hipparcos catalog lists the parallax of HD~98800 as 21.43 $\pm$
2.86 mas, corresponding to a system distance of 46.7 $\pm$ 6.2 pc
\citep{HIP1997}.  The distance determination to HD~98800~B based on
our orbital solution is 42.2 $\pm$ 4.7 pc, corresponding to an
orbital parallax of 23.7 $\pm$ 2.6 mas, consistent with the
Hipparcos result at 9.5\% and 0.6-sigma.

At the distance of HD~98800 neither of the B subsystem components are
significantly resolved by the KI K-band fringe spacing, and we must
resort to model diameters for the components.  We have estimated the
HD~98800~B component apparent diameters through spectral energy
distribution modeling.  We find apparent diameters of 0.238 $\pm$
0.017 and 0.188 $\pm$ 0.014 mas for the Ba and Bb components
respectively (details of the spectral energy distribution modeling are
given in \S\ref{sec:SED}).  With our system distance estimate these
estimated diameters correspond to physical radii of 1.09 $\pm$ 0.14
and 0.85 $\pm$ 0.11 R$_\sun$, and (combined with the mass estimates)
log surface gravities of 4.21 $\pm$ 0.12 and 4.34 $\pm$ 0.12 for the
Ba and Bb components respectively.

\begin{deluxetable}{ccc}
\tabletypesize{\small} \tablecolumns{3} \tablewidth{0pc}

\tablecaption{Physical Parameters for HD~98800~B.  Summarized here are
the physical parameters for the HD~98800~B subsystem as derived
from the ``Joint-Fit'' solution orbital parameters in Table
\ref{tab:orbit}, and SED modeling.
\label{tab:physics}
}

\tablehead{
\colhead{Physical}   & \colhead{Ba}         & \colhead{Bb} \\
\colhead{Parameter}  & \colhead{Component}  & \colhead{Component}
}

\startdata
a (10$^{-1}$ AU)     & 4.47 $\pm$ 0.13      & 5.36 $\pm$ 0.13    \\
Mass (M$_{\sun}$)    & 0.699 $\pm$ 0.064    & 0.582 $\pm$ 0.051  \\
\cline{2-3}
System Distance (pc) & \multicolumn{2}{c}{42.2 $\pm$ 4.7} \\
$\pi_{orb}$ (mas)    & \multicolumn{2}{c}{23.7 $\pm$ 2.6} \\
\cline{2-3}
T$_{eff}$ (K)        & 4200 $\pm$ 150       & 4000 $\pm$ 150   \\
Model Diameter (mas) & 0.239 $\pm$ 0.017    & 0.188 $\pm$ 0.014  \\
Bolometric Flux (10$^{-9}$erg cm$^{-2}$ s$^{-1}$)
                     & 5.96 $\pm$ 0.28   & 3.00 $\pm$ 0.15 \\
Luminosity (L$_{\sun}$)
                     & 0.330 $\pm$ 0.075 & 0.167 $\pm$ 0.038 \\
Radius (R$_{\sun}$)  & 1.09 $\pm$ 0.14   & 0.85 $\pm$ 0.11 \\
$\log g$             & 4.21 $\pm$ 0.12   & 4.34 $\pm$ 0.12 \\

M$_K$ (mag)          & 3.80 $\pm$ 0.25   & 4.38  $\pm$ 0.25  \\
M$_V$ (mag)          & 6.91 $\pm$ 0.26   & 8.02  $\pm$ 0.27  \\
$V$-$K$ (mag)        & 3.11 $\pm$ 0.12   & 3.61  $\pm$ 0.14  \\
\enddata

\end{deluxetable}

\subsection{Spectral Energy Distribution Modeling}
\label{sec:SED}
Because interferometric observations potentially resolve the stellar
components in a binary system, we always construct spectral energy
distribution (SED) models for binary systems to estimate a priori
apparent diameters.  For HD~98800~B SED modeling results are
particularly interesting in light of a conjecture by
\citet{Tokovinin1999} that the putative B circumbinary disk (P2001)
obscures our view [of the B sub]system to explain the irregular
low-level photometric variability observed by Hipparcos (S98).

\begin{figure}[t]
\includegraphics[angle=-90,width=14.5cm]{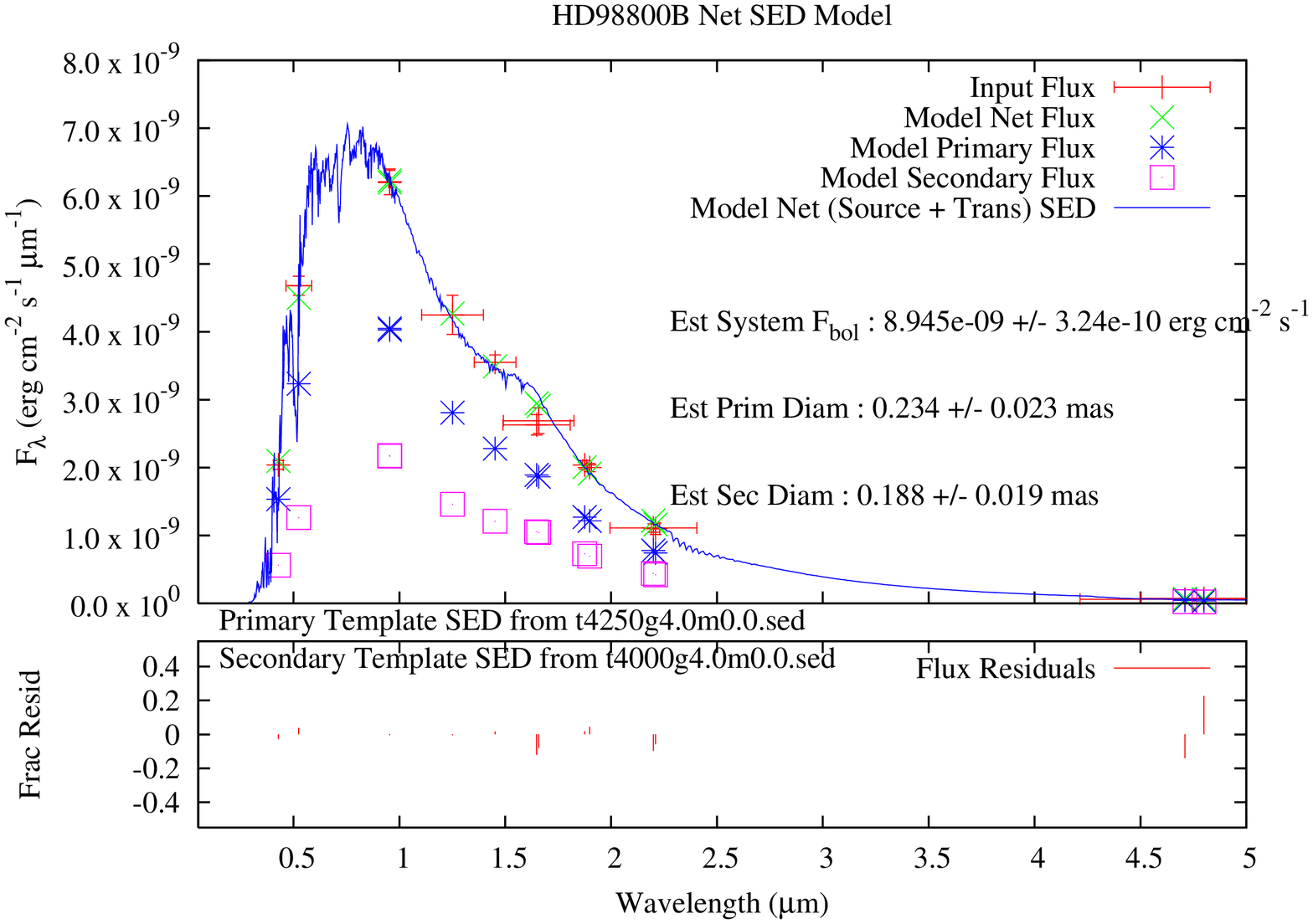}\\
\includegraphics[angle=-90,width=15.5cm]{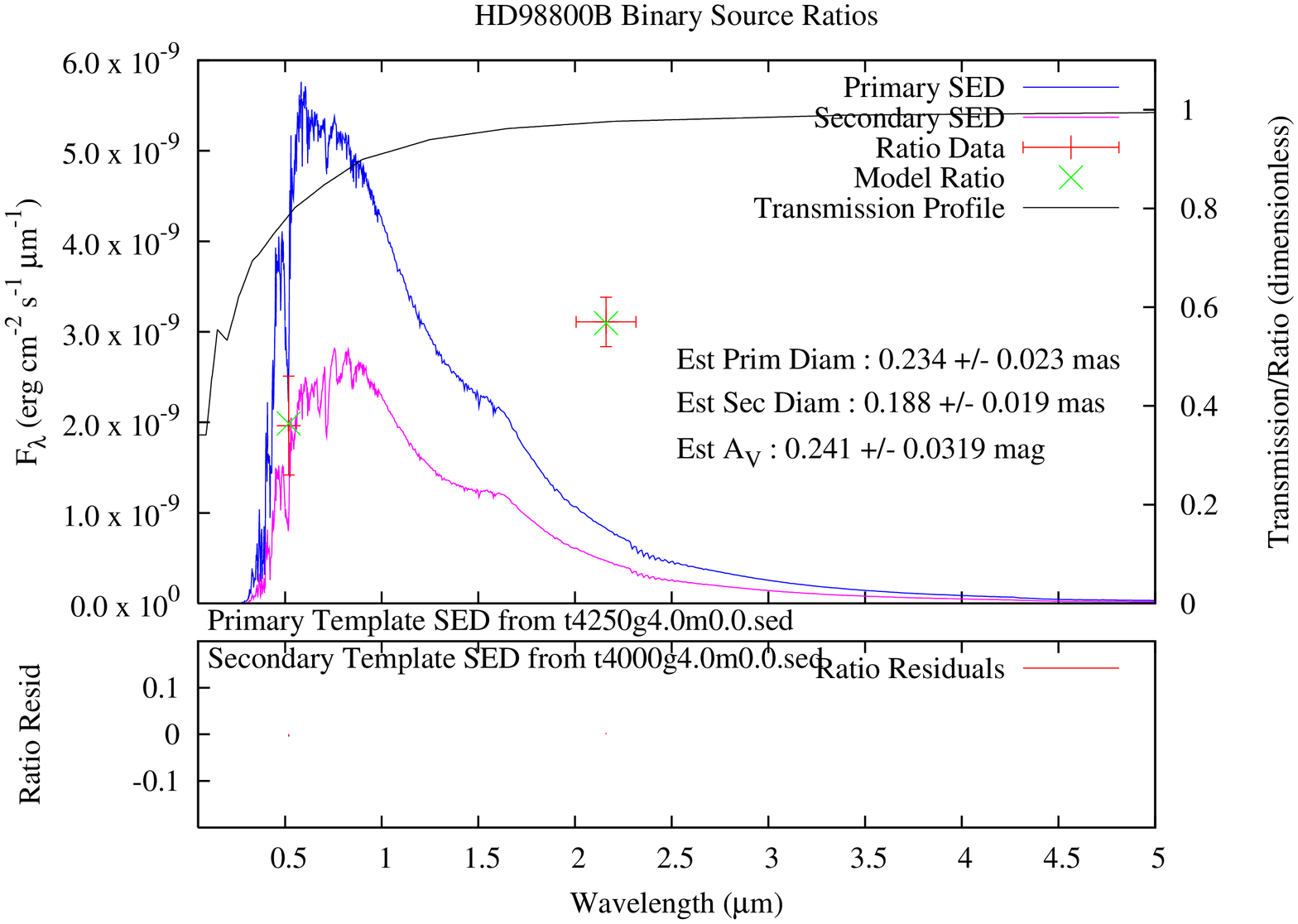}\\
\caption{Sample Spectral Energy Distribution Model for HD~98800~B.
Here SED templates from \citet{Lejeune1997,Lejeune1998} have been used
to simultaneously model published flux measurements (summarized in
P2001 Table 2; top) and component flux ratio estimates (Table
\ref{tab:orbit}; bottom) from our KI V$^2$ and T95 spectroscopic
measurements.
\label{fig:HD98800B_SED}}
\end{figure}

P2001
summarizes the body of A/B-resolved flux measurements in the HD~98800
system, and has modeled SEDs of the A and B subsystems with Kurucz
templates.  To the P2001 flux data 
have we add Bb/Ba flux ratio estimates based on our visibility and
spectroscopic observations.  We estimate a Bb/Ba flux ratio of 0.36
$\pm$ 0.05 at 519 nm in the CfA spectroscopy based on the original
spectra.  This estimate is different than the original T95 value -- it
includes a correction related to the different normalization of the
template spectra used in T95, which was overlooked in the original
analysis.  At 2.2 $\mu$m the flux ratio is 0.569 $\pm$ 0.024 from the
KI visibility data (Table \ref{tab:orbit}).  Using a custom
two-component SED modeling code, we have modeled the B subsystem flux
and ratios using solar and sub-solar abundance SED templates from
\cite{KuruczModels}, \citet{Lejeune1997,Lejeune1998}, and
\citet{Pickles1998}, and compared the flux and ratio data with a large
grid of SED templates for each component ranging in temperature
(spectral type) from 3500 -- 5000 K (M3 -- K2).  Good fits were found
with both Lejeune and Pickles SED templates, and the results
consistently preferred temperature of 4200 $\pm$ 150 K for Ba and 4000
$\pm$ 150 K for Bb; these values are consistent with results from
other studies (e.g.~S98, P2001).
Figure \ref{fig:HD98800B_SED} depicts the best-fit results from
Lejeune template comparisons.  Component temperatures, angular
diameters, and related quantities in Table~\ref{tab:physics} are
composite values from Lejeune and Pickles SED model results.

Notable in the SED solutions is the necessity to include a small
amount of extinction (A$_V$ $\sim$ 0.3 $\pm$ 0.05) to model the
observed B subsystem flux distribution.  S98
reached a similar conclusion in their analysis of Li-line spectral
observations (including this finding as part of their favored
hypothesis for the B component parameters).  It is tempting to take
this extinction as corroboration of Tokovinin's conjecture that B
circumbinary material lies along the line of sight to the stellar
components.  As a cross-check of this hypothesis we ran a similar
series of SED models for the A component (the Ab component was taken
to contribute zero flux), and found that no extinction was necessary
to fit the A flux distribution.  Based on our SED modeling it appears
that there is a source of extinction along our line of sight to B that
is absent to A.

P2001 has argued that the putative B circumbinary disk is likely
coplanar with the B orbital plane.  A simple coplanar disk would
require an aspect ratio or flaring of at least $\tan(90 - 67) \approx
0.42$ for such a disk to be the source of this extinction.
Alternatively, it seems likely that a B circumbinary disk would be
disrupted by the A -- B orbit, so the presumption of simple disk
co-planarity with the B orbit may not be well founded.  It seems direct
resolution and mapping of the B circumbinary disk will be critical to
understand the B circumbinary material distribution in the highly
dynamic HD~98800 system.

\section{Discussion}
\label{sec:discussion}

\subsection{Comparison with stellar evolution models}
\label{sec:modelcomp}

\citet{Hillenbrand2004} have shown that PMS models are in good
agreement with dynamically determined masses above 1.2 M$_\sun$.
However, below 1.2 M$_\sun$ the existing models do a poorer job of
matching observed component properties, tending to predict hotter and
more luminous stars for a given mass.  Of course, it is exactly this
kind of apparent model/observation discrepancy that makes HD~98800
such a compelling system for careful study.  While the orbit results
presented here are preliminary, it is still interesting to make some
initial comparisons between our observations and PMS evolutionary
models.

\begin{figure}[t]
\epsscale{1.1}
\plottwo{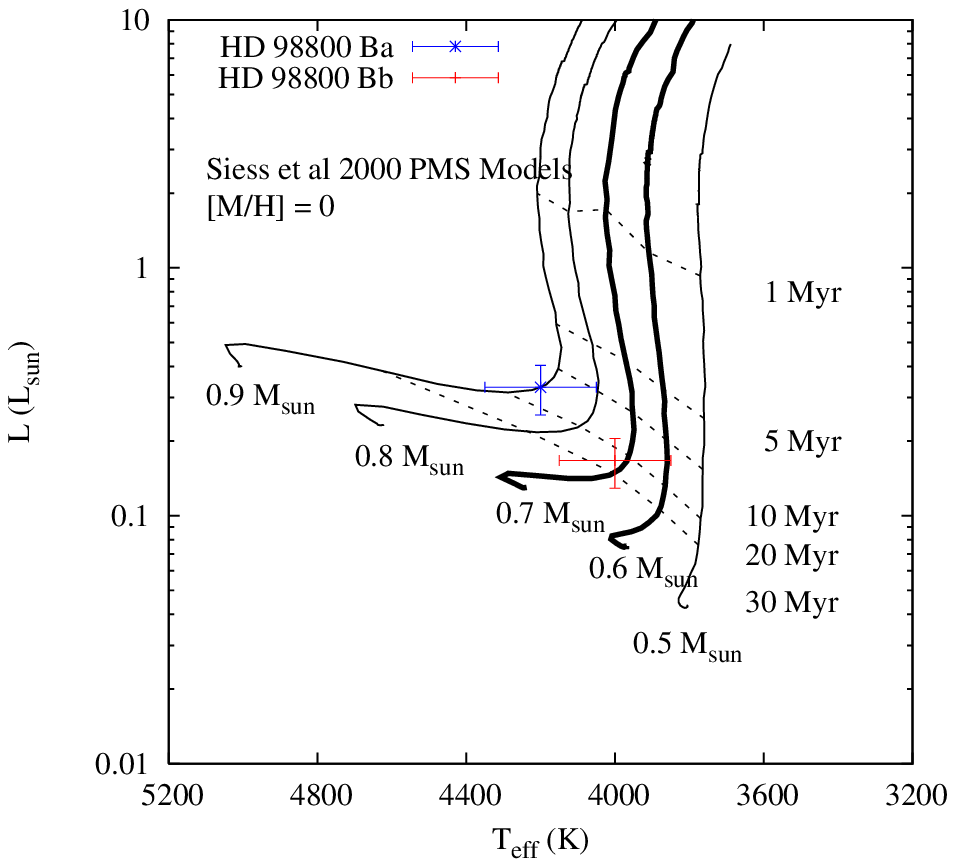}{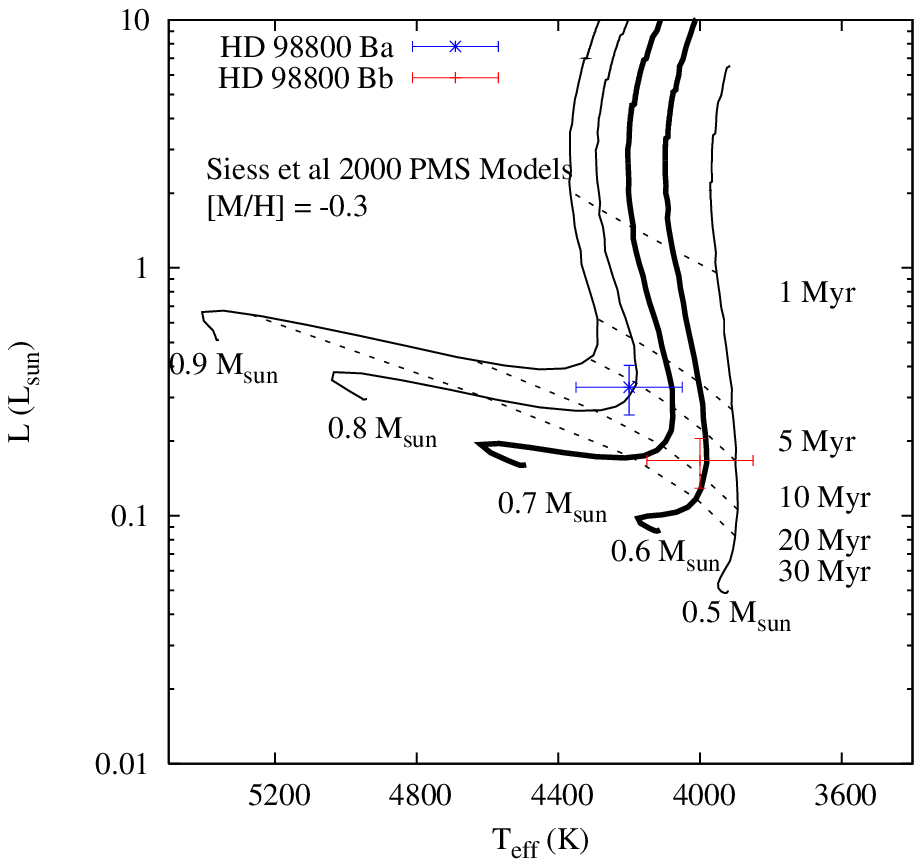}\\
\plottwo{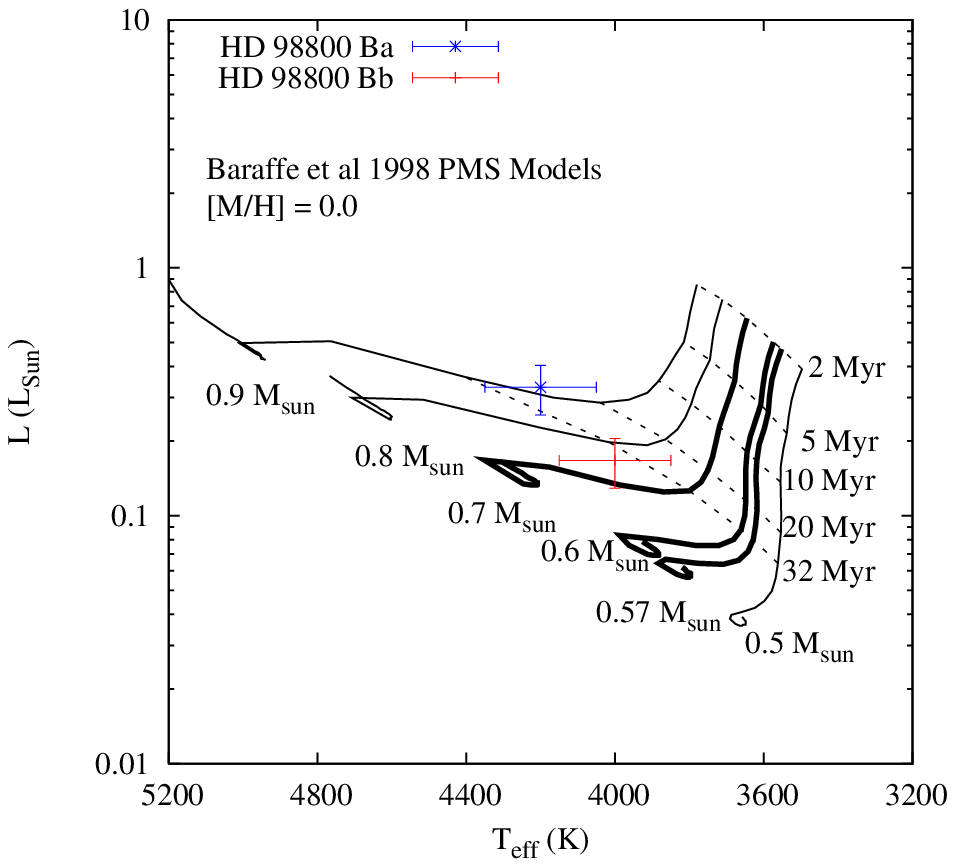}{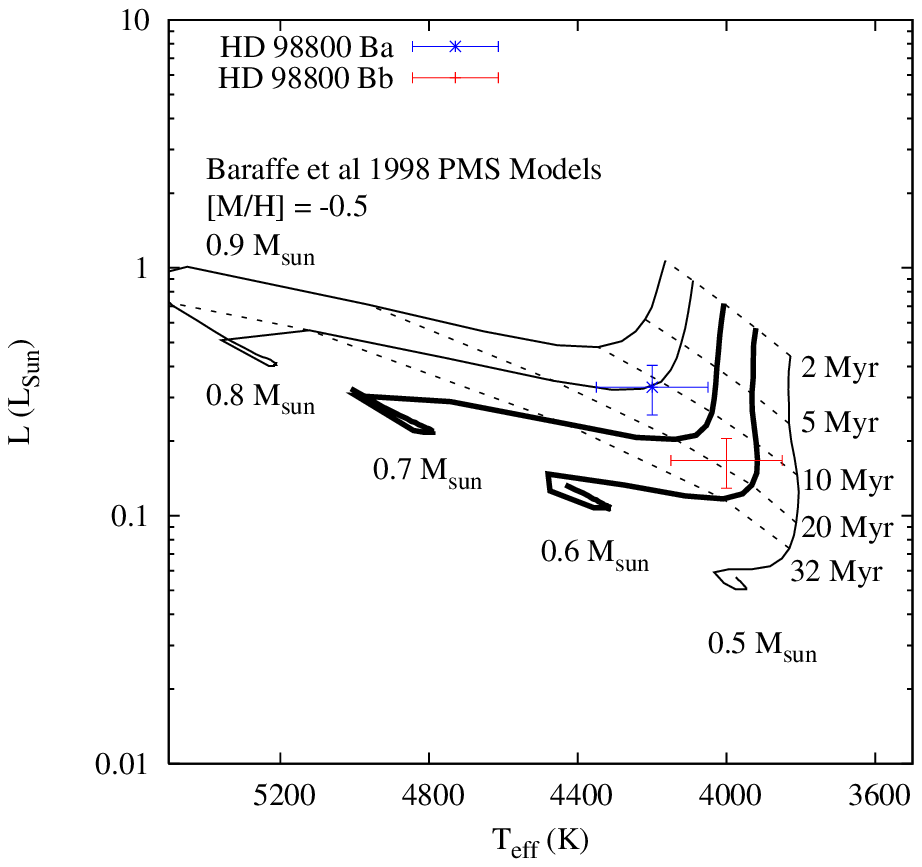}\\
\caption{HD~98800~B Components Compared With PMS Models.  Here we show
the HD~98800~B components in luminosity/T$_{eff}$ space, and PMS
evolutionary tracks by \citet{Siess2000} (top panels) and
\citet{Baraffe1998} (bottom panels).  As found by P2001,
solar abundance models (left panels) suggest higher masses for the
HD~98800~B than our orbit yields, however a lower abundance ([M/H] =
-0.3 -- -0.5, right panels) for the components brings the inferred
masses and radiative properties into better agreement with model
predictions from both families.  On all panels bold lines emphasize
mass tracks bracketing component mass values implied by the orbit
model (Table~\ref{tab:physics}), and isochrones at spanning a range of
ages between 1 and 32 Myrs are given.
\label{fig:modelCompare}}
\end{figure}

Our inferred radiometric parameters for the HD~98800~B components
(Table~\ref{tab:physics}) are in good agreement to those found in
previous work (S98, P2001),
so comparisons with PMS models largely go along similar lines.
Figure~\ref{fig:modelCompare} shows the position of the B subsystem
components in luminosity/T$_{eff}$ space, along with PMS evolutionary
models from \citet{Siess2000} and \citet{Baraffe1998}.  Model
evolutionary tracks for masses that bracket the component masses
inferred from our orbit model are emphasized in the figure.  As found
by P2001,
for solar metalicity (Figure~\ref{fig:modelCompare} left panels) the
radiative properties of the Ba and Bb components predict component
masses that are significantly higher than our orbit model would
indicate. This naturally led P2001 to infer a B orbital inclination
($\sim$ 58$^\circ$) that is similar but slightly lower than the orbit
model presented here.  Both \citet{Siess2000} and \citet{Baraffe1998}
models predict slightly cooler and less luminous components at the low
masses indicated by our orbit model.

The apparent model/observation discrepancy is reduced if we consider
the possibility that elemental abundances for HD~98800 components are
sub-solar.  S98
attempted an abundance estimate for HD~98800, and argue for solar
abundance with an uncertainty of $\sim$ 0.2 dex.
Figure~\ref{fig:modelCompare} right panel shows the same comparison of
component properties with \citet{Siess2000} models at [M/H] = -0.3
(allowed by S98), and \citet{Baraffe1998} models at [M/H] = -0.5
(probably not allowed by S98).
The match between the lower abundance models at our inferred component
masses and radiometric properties is improved, with both components
matching the relevant mass tracks within the temperature error bars
for both the lower abundance \citet{Siess2000} and \citet{Baraffe1998}
models.  From superimposed isochrones in Figure~\ref{fig:modelCompare}
we infer the HD~98800 age is in the 8 -- 20 MYr range, consistent with
previous findings (S98).

\clearpage

\clearpage

\subsection{Conclusions and Future Work}
\label{sec:conclusions}
The preliminary orbit presented here for HD~98800~B does an excellent
job of matching up with other empirical constraints on the system: it
independently phases properly with the T95 RV solution, and it yields
a distance estimate that agrees well with the Hipparcos determination.
The inclination of the orbit model raises interesting questions
concerning the nature of the putative B circumbinary disk.  A planar
disk and strict orbit/disk co-planarity would imply a thick aspect
ratio or flaring for the disk ($\approx$ 0.4) if it produces the
inferred extinction along the line of sight to the B components,
however perturbations to the disk from the outer A-B orbit seem
likely.  Further, the implied physical semi-major axis of the B orbit
(Table \ref{tab:physics}; 0.98 $\pm$ 0.02 AU) fits within previous
estimates of the inner disk radius of 1.5 -- 2 AU (S98, P2001), but it
is unclear whether the implied separation is consistent with dynamical
modeling of such circumbinary disks
\citep[e.g.][]{Artymowicz1994,Pichardo2005}.

The orbit model presented here implies component dynamical masses
accurate to approximately 8\%; with such mass determinations we can
not yet provide critical tests for low-mass PMS stellar models.  Prima
facie comparisons with PMS models from \citet{Baraffe1998} and
\citet{Siess2000} suggest that the components are hotter and more
luminous than predicted by the models for solar abundance, and
considering the possibility of slightly sub-solar abundance brings the
inferred component parameters into better agreement with the model
predictions.  However at present the significant dynamical mass and
effective temperature errors do not yet allow for definitive
conclusions.  Narrowly it is important to establish the abundance of
HD~98800 to properly test and refine PMS models.  More broadly it is
intriguing to consider the possibility that stars as young as TWA
members may have significantly sub-solar abundances.

The B subsystem orbit presented here is preliminary, and we are
continuing observation of the system interferometrically and
spectroscopically to refine the orbit model, and resulting component
properties and system distance.  The apparent inclination for B is
such that additional observations are likely to result in
significantly refined component parameter estimates.  A forthcoming
publication will provide a more detailed comparison between the B
component parameters and PMS models based on a refined B subsystem
orbit model.  Further, we have also observed HD~98800~A with KI on two
epochs, and the V$^2$ data show clear signs of resolving that
subsystem as well -- the first time the Ab component has been directly
detected.  We will continue to observe A to determine its orbit and
component properties as well.

\acknowledgements 

The authors wish to thank the anonymous reviewer for the many
thoughtful comments that have greatly improved this manuscript.

Part of this work was performed at the Michelson Science Center (MSC),
California Institute of Technology under contract with the National
Aeronautics and Space Administration (NASA).

Some of the data presented herein were obtained at the W.M.~Keck
Observatory, which is operated as a scientific partnership among the
California Institute of Technology, the University of California and
the NASA.  The Observatory was made possible by the generous financial
support of the W.M.~Keck Foundation.  We gratefully acknowledge the
support of personnel at the Jet Propulsion Laboratory, W.M.~Keck
Observatory, and the MSC in obtaining KI observations of HD~98800.
The authors wish to recognize and acknowledge the very significant
cultural role and reverence that the summit of Mauna Kea has always
had within the indigenous Hawaiian community.  We are most fortunate
to have the opportunity to conduct observations from this mountain.

GT acknowledges partial support from NASA's MASSIF SIM Key Project
(BLF57-04) and NSF grant AST-0406183.

This research has made use of services of the MSC at the California
Institute of Technology; the SIMBAD database, operated at CDS,
Strasbourg, France; of NASA's Astrophysics Data System Abstract
Service; and of data products from the Two Micron All Sky Survey,
which is a joint project of the University of Massachusetts and the
Infrared Processing and Analysis Center, funded by NASA and the
National Science Foundation.

\clearpage

\end{document}